\documentclass[11pt, letter]{article}
\usepackage{amssymb,amsmath,array}
\usepackage{fullpage}
\usepackage{times}
\usepackage{helvet}
\usepackage{courier}

\usepackage{xspace}

\usepackage{algorithmic}	
\usepackage{algorithm}		
\usepackage{amsmath,multirow,amssymb,bm}

\usepackage{hyperref}

\newcommand{\RR}{\mathbb{R}}

\newcommand{\AB}{\mathcal{A}}
\newcommand{\KW}{\mathcal{K}}
\newcommand{\SM}{\mathcal{S}}

\newcommand{\NET}{\mathcal{N}}

\usepackage{xspace}
\newcommand{\ftext}{\textsf{FastText}\xspace}
\newcommand{\sysname}{\textsf{MOLIERE}\xspace}
\newcommand{\tmine}{\textsf{ToPMine}\xspace}

\newcommand{\arrowsmith}{\textsf{Arrowsmith}\xspace}

\usepackage{changes}
\definechangesauthor[color=red]{is}
\definechangesauthor[color=blue]{js}

\usepackage{graphicx}
\usepackage{calc}
\newlength{\depthofsumsign}
\newlength{\totalheightofsumsign}
\newlength{\heightanddepthofargument}

\makeatletter
\newcommand*{\DivideLengths}[2]{%
  \strip@pt\dimexpr\number\numexpr\number\dimexpr#1\relax*65536/\number\dimexpr#2\relax\relax sp\relax
}
\makeatother

\begin{document}
\title{\sysname: Automatic Biomedical Hypothesis Generation System}

\author{Justin Sybrandt$^1$, Michael Shtutman$^2$, Ilya Safro$^1$
%
%
\vspace{.3cm}\\
%
1- Clemson University, School of Computing, Clemson SC, USA
%
\vspace{.1cm}\\
2- University of South Carolina, Drug Discovery and Biomedical Sciences, Columbia SC, USA
}

\maketitle

\begin{abstract}
Hypothesis generation is becoming a crucial time-saving technique which allows biomedical researchers to quickly discover implicit connections between important concepts. Typically, these systems operate on domain-specific fractions of public medical data. \sysname, in contrast, utilizes information from over 24.5 million documents. At the heart of our approach lies a multi-modal and multi-relational network of biomedical objects extracted from several heterogeneous datasets from the National Center for Biotechnology Information (NCBI). These objects include but are not limited to scientific papers, keywords, genes, proteins, diseases, and diagnoses. We model hypotheses using Latent Dirichlet Allocation applied on abstracts found near shortest paths discovered within this network, and demonstrate the effectiveness of \sysname by performing hypothesis generation on historical data. Our network, implementation, and resulting data are all publicly available for the broad scientific community. 

\end{abstract}

\section{Introduction}



Vast amounts of biomedical information accumulate in modern databases such as MEDLINE \cite{Medicine2016}, which currently contains the bibliographic data of over 24.5 million medical papers. These ever-growing datasets impose a great difficulty on researchers trying to survey and evaluate new information in the existing biomedical literature, even when advanced ranking methods are applied. On the one hand, the vast quantity and diversity of available data has inspired many scientific breakthroughs. On the other hand, as the set of searchable information continues to grow, it becomes impossible for human researchers to query and understand all of the data relevant to a domain of interest.

In 1986 Swanson hypothesized that novel discoveries could be found by carefully studying the existing body of scientific research \cite{swanson1986undiscovered}. Since then, many groups have attempted to mine the wealth of public knowledge. Efforts such as Swanson's own \arrowsmith generate hypotheses by finding concepts which implicitly link two queried keywords. His method and others are discussed at length in Section \ref{relatedWork}. Ideally, an effective hypothesis generation system greatly increases the productivity of researchers. For example, imagine that a medical doctor believed that stem cells could be used to repair the damaged neural pathways of stroke victims (as some did in 2014 \cite{Hao2014}). If no existing research directly linked stem cells to stroke victims, this doctor would typically have no choice but to follow his/her intuition. Hypothesis generation allows this researcher to quickly learn the likelihood of such a connection by simply running a query. Our hypothetical doctor may query the topics \textit{stem cells} and \textit{stroke} for example. If the system returned topics such as \textit{paralysis} then not only would the doctor's intuition be validated, but he/she would be more likely to invest in exploring such a connection. In this manner, an intelligent hypothesis generation system can increase the likelihood that a researcher's study yields usable new findings. 

\subsection{Our Contribution}

We introduce a deployed system, \sysname \cite{oursystem2017}, with the goal of generating more usable results than previously proposed hypothesis generation systems. We develop a novel method for constructing a large network of public knowledge and devise a query process which produces human readable text highlighting the relationships present between nodes.

To the best of our knowledge, \sysname is the first hypothesis generation system to utilize the entire MEDLINE data set. By using state-of-the-art tools, such as \tmine \cite{el2014scalable} and \ftext \cite{Bojanowski2016}, we are able to find novel hypotheses without restricting the domain of our knowledge network or the resulting vocabulary when creating topics. As a result, \sysname is more generalized and yet still capable of identifying useful hypotheses. 

We provide our network and findings online for others in the scientific community \cite{oursystem2017}. Additionally, to aid interested biomedical researchers, we supply an online service where users can request specific query results at \url{http://jsybran.people.clemson.edu/mForm.php}. 
Furthermore, \sysname is entirely open-source in order to facilitate similar projects. See \url{https://github.com/JSybrandt/MOLIERE} for the code needed to generate and query the \sysname knowledge network.

In the following paper we describe our process for creating and querying a large knowledge network built from MEDLINE and other NCBI data sources. We use natural language processing methods, such as Latent Dirichlet Allocation (LDA) \cite{blei2003latent} and topical phrase mining \cite{el2014scalable}, along with other data mining techniques to conceptually link together abstracts and biomedical objects (such as biomedical keywords and $n$-grams) in order to form our network. Using this network we can run shortest path queries to discover a pathway between two concepts which are non-trivially connected. We then find clouds of documents around these pathways which contain knowledge representative of the path as a whole. PLDA+, a scalable implementation of LDA \cite{liu2011plda+}, allows us to quickly find topic models in these clouds. Unlike similar systems, we do not restrict PLDA+ to any set vocabulary. Instead, by using topical phrase mining, we identify meaningful n-grams in order to improve the performance, flexibility, and understandability of our LDA models. These models result in both quantitative and qualitative connections which human researchers can use to inform their decision making.

We evaluate our system by running queries on historical data in order to discover landmark findings. 
For example, using data published on or before 2009, we find strong evidence that the protein Dead Box RNA Helicase 3 (DDX3) can be applied to treat cancer. We also verify the ability of \sysname to make predictions similar to previous systems with restricted LDA \cite{wang2011finding}.

\subsection{Our Method in Summary}

We focus on the domain of medicine because of the large wealth of public information provided by the National Library of Medicine (NLM). MEDLINE is a database containing over 24.5 million references to medical publications dating all the way back to the late 1800s \cite{Medicine2016}. Over 23 million of these references include the paper's title and abstract text. In addition to MEDLINE, the NLM also maintains the Unified Medical Language System (UMLS) which is comprised of three main resources: the metathesaurus, the semantic network, and the SPECIALIST natural language processing (NLP) tools. These resources, along with the rest of our data, are described in section \ref{data}.

Our knowledge base starts as XML files provided by MEDLINE, from which we extract each publication's title, document ID, and abstract text. We first process these results with the SPECIALIST NLP toolset. The result is a corpus of text which has standardized spellings (for example ``colour'' becomes ``color''), no stop words (including medical specific stop words such as \textit{Not Otherwise Specified (NOS)}), and other characteristics which improve later algorithms on this corpus. Then we use \tmine to identify multi-word phrases from that corpus such as ``asthma attack,"  allowing us to treat phrases as single tokens \cite{el2014scalable}. Next, we send the corpus through \ftext, the most recent word2vec implementation, which maps each unique token in the corpus to a vector  \cite{mikolov2013efficient}. 
We can then fit a centroid to each publication and use the Fast Library for Approximate Nearest Neighbors (FLANN) to generate a nearest neighbors graph \cite{muja2014scalable}. The result is a network of MEDLINE papers, each of which are connected to other papers sharing a similar topic. This network, combined with the UMLS metathesaurus and semantic network, constitutes our full knowledge base. The network construction process is described in greater detail in Section \ref{sec:networkConstruction}.

With our network, a researcher can query for the connections between two keywords. We find the shortest path between the two keywords in the knowledge network, and extend this  path to identify a significant set of related abstracts. This subset contains many documents which, due to our network construction process, all share common topics. We perform topic modeling on these documents using  PLDA+ \cite{liu2011plda+}. The result is a set of plain text topics which represent different concepts which likely connect the two queried keywords. More information about the query process is detailed in Section \ref{sec:queryProcess}.

We use landmark historical findings in order to validate our methods. For example, we show the implicit link between Venlafaxine and HTR1A, and the involvement of DDX3 on Wnt signaling. These queries and results are detailed in Section \ref{experiments}.
In Sections \ref{sec:deploymentChallenges} and \ref{sec:lessonsLearned} we discuss challenges and  open research questions we have uncovered during our work.

\subsection{Related Work} \label{relatedWork}

The study and exploration of undiscovered public knowledge began in 1986 with Swanson's landmark paper \cite{swanson1986undiscovered}. Swanson hypothesized that fragments of information from the set of public knowledge could be connected in such a way as to shed light on new discoveries. With this idea, Swanson continued his research to develop \arrowsmith, a text-based search application meant to help doctors make connections from within the MEDLINE data set \cite{smalheiser1998using,swanson1998link,swanson1999implicit}. To use \arrowsmith, researchers supply two UMLS keywords which are used to find two sets of abstracts, $A$ and $C$. The system then attempts to find a set $B \approx A \cap C$. Assuming sets $A$ and $C$ do not overlap initially, implicit textual links are used to expand both sets until some sizable set B is discovered. The experimental process was computationally expensive, and queries were typically run on a subset of the MEDLINE data set (according 
to \cite{swanson1999implicit} around 1,000 documents). 

Spangler has also been a driving force in the field of hypothesis generation and mining undiscovered public knowledge. His textbook \cite{spangler2015accelerating} details many text mining techniques as well as an example application related to hypothesis generation in the MEDLINE data set. His research in this field has focused on p53 kinases and how these undiscovered interactions might aid drug designers \cite{spangler2014automated,spangler2015accelerating}. His method leverages unstructured text mining techniques to identify a network entities and relationships from medical text. Our work differs from this paradigm by utilizing the structured UMLS keywords, their known connections, and mined phrases. We do, however, rely on similar unstructured text mining techniques, such as \ftext and FLANN, to make implicit connections between the abstracts.

Rzhetsky and Evans notice that current information gathering methods struggle to keep up with the growing wealth of forgotten and hard to find information \cite{evans2011advancing}. Their work in the field of hypothesis generation has included a study on the assumptions made when constructing biomedical models \cite{divoli2011conflicting} and  digital representations of hypothesis \cite{soldatova2011representation}.

Divoli et al. analyze the assumptions made in medical research \cite{divoli2011conflicting}. They note that scientists often reach contradictory conclusions due to differences in each person's underly assumptions. The study in \cite{divoli2011conflicting} highlights the variance of these preconceptions by surveying medical researchers on the topic of cancer metastasis. Surprisingly, 27 of the 28 researchers surveyed disagree with the textbook process of cancer metastasis. When asked to provide the ``correct'' metastasis scenario, none of the surveyed scientists agree. Divoli's study highlights a major problem for hypothesis generation. Scientists often disagree, even in published literature. Therefore, a hypothesis generation system must be able to produce reliable results from a set of contradicting information.

In \cite{soldatova2011representation}, Soldatova and Rzhetsky describe a standardized way to represent scientific hypotheses. By creating a formal and machine readable standard, they envision a collection of hypotheses which clearly describes the full spectrum of existing theories on a given topic. Soldatova and Rzhetsky extend existing approaches by representing hypotheses as logical statements which can be interpreted by \textit{Adam}, a robot scientist capable of starting one thousand experiments a day. Adam is successful, in part, because they model hypotheses as an ontology which allows for Bayesian inference to govern the likelihood of a specific hypothesis being correct.

DiseaseConnect, an online system that allows researchers to query for concepts intersecting two keywords, is a notable contribution to hypothesis generation \cite{liu2014diseaseconnect}. This system, proposed by Liu et al., is similar to both our system and \arrowsmith \cite{smalheiser2009arrowsmith} in its focus on UMLS keywords and MEDLINE literature mining. 
Unlike our system, Liu et al. restrict DiseaseConnect to simply 3 of the 130 semantic types. They supplement this subset with concepts from the OMIM \cite{goh2007human} and GWAS \cite{barrenas2009network} databases, two genome specific data sets.  Still, their network size is approximately 10\% of the size of \sysname.
DiseaseConnect uses its network to identify diseases which can be grouped by their molecular mechanisms rather than symptoms. The process of finding these clusters depends on the relationships between different types of entities present in the DiseaseConnect network. Users can view sub-networks relevant to their query online and related entities are displayed alongside the network visualization.

Barab{\'a}si et al. improve upon the network analytic approach to understand biomedical data in both their work on the disease network \cite{goh2007human} as well as their more generalized symptoms-disease network  \cite{zhou2014human}. 
In the former \cite{goh2007human}, the authors  construct a bipartite network of disease phonemes and genomes to which they refer to as the \textit{Diseasome}. Their inspiration is an observation that genes which are related to similar disorders are likely to be related themselves. They use the \textit{Diseasome} to create two projected networks, the human disease network (HDN), and the Disease Gene Network (DGN). 
In the latter \cite{zhou2014human}, they construct a more generalized human symptoms disease network (HSDN) by using both UMLS keywords and bibliographic data. HSDN consists of data collected from a subset of MEDLINE consisting of only abstracts which contained at least one disease as well as one symptom, a subset consisting of approximately 850,000 records. From this set, Goh et al. calculated keyword co-occurrence statistics in order to build their network. They validate their approach using 1,000 randomly selected MEDLINE documents and, with the help of medical experts, manually confirm that the relationship described in a document is reflected meaningfully in HSDN. Ultimately, Goh et al. find strong correlations between the symptoms and genes shared by  common  diseases. 

Bio-LDA is a modification of LDA which limits the set of keywords to the set present in UMLS \cite{wang2011finding}. This reduction improves the meaning and readability of topics generated by LDA. Wang et al. also show in this work that their method can imply connections between keywords which do not show up in the same document. For example, they note that Venlafaxine and HTR1A both appear in the same topic even though both do not appear in the same abstract. We explore and repeat these findings in Section \ref{sec:venToHTR}.

\subsection{Related and Incorporated Technologies}
\label{sec:relTech}

\textbf{\ftext} is the most recent implementation of word2vec from Milkolov et al. \cite{mikolov2013efficient,mikolov2013distributed, joulin2016bag,Bojanowski2016}. Word2vec is a method which utilizes the skip-gram model to identify the relationships between words by analyzing word usage patterns. This process maps plain text words into a high dimensional vector space for use in data mining applications. Similar words are often grouped together, and the distances between words can reveal relationships. For example, the distance between the words ``Man" and ``Woman" is approximately the same as the distance between ``King" and ``Queen". \ftext improves upon this idea by leveraging sub-strings in long rarely occurring words. 

\textbf{\tmine}, a project from El-Kishky et al., is focused on discovering multi-word phrases from a large corpus of text \cite{griffiths2004finding}. This project intelligently groups unigrams together to create n-gram phrases for later use in text mining algorithms. By using a bag-of-words topic model, \tmine groups unigrams based on their co-occurrence rate as well as their topical similarity using a process they call Phrase LDA.

\textbf{Latent Dirichlet Allocation} \cite{blei2003latent} is the most common topic modeling process and PLDA+ is a scalable implementation of this algorithm \cite{griffiths2004finding,liu2011plda+}. Developed by Zhiyuan Liu et al., PLDA+ quickly identifies groups of words and phrases which all relate to a similar concept. Although it is an open research question as to how best to interpret these results, simple qualitative analysis allows for ``ballpark" estimations. For instance, it may take a medical researcher to wholly understand the topics generated from abstracts related to two keywords, but anyone can identify that all words related to a concept of interest occur in the same topic. Results like this, show that LDA has distinguished the presence of a concept in a body of text.

\section{Knowledge Network Construction} \label{sec:networkConstruction}

\begin{figure}
\includegraphics[width=\linewidth]{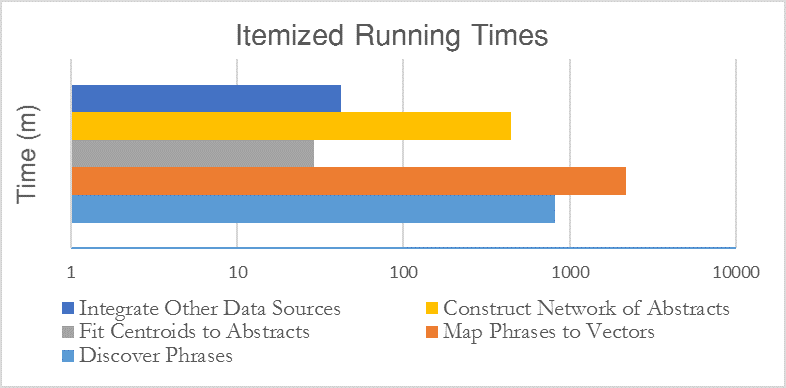}
\caption{Running times of each network construction phase. All phases run on a single node described in section \ref{sec:expirementalSetup}. Not shown: Initial text processing which was handled by a large array of small nodes.}
\label{fig:runTime}
\end{figure}

In order to discover hypotheses we construct a large weighted multi-layered network of biomedical objects extracted from NLM  data sets. Using this network, we run shortest-centroid-path queries (see Section \ref{sec:queryProcess}) whose results serve as an input for hypothesis mining. The wall clock time needed to complete this network construction pipeline is depicted in Figure \ref{fig:runTime} (see details in Section \ref{sec:expirementalSetup} ).  Omitted from this figure is the time spent preprocessing the initial abstract text due to its embarrassingly parallel nature.

\subsection{Data Sources} \label{data}
The NLM maintains multiple databases of medical information which are the main source of our data. This includes MEDLINE \cite{Medicine2016}, a source containing the metadata of approximately 24.5 million medical publications since the late 1800's. Most of these MEDLINE records include a paper's title, authors, publication date, and abstract text. 

In addition to MEDLINE, the NLM maintains UMLS \cite{MedicineUS2009}, which in turn provides the metathesaurus as well as a semantic network. 
The metathesaurus contains two million keywords along with all known synonyms (referred to as ``atoms'') used in medical text. For example, the keyword ``RNA'' has many different synonyms  such as ``Ribonucleinicum acidum'', ``Ribonucleic Acid'', and ``Gene Products, RNA'' to name a few.
These metathesaurus keywords form a network comprised of multi-typed edges. For example, an edge may represent a \textit{parent - child} or a \textit{boarder concept - narrower concept} relationship.  RNA has connections to terms such as ``Nucleic Acids'' and ``DNA Synthesizers''.  
Lastly, each keyword holds a reference to an object in the semantic network. RNA is an instance of the ``Nucleic Acid, Nucleoside, or Nucleotide'' semantic type.

The UMLS semantic network is comprised of approximately 130 semantic types and is connected in a similar manner as the metathesaurus. For example, the semantic type ``Drug Delivery Device'' has an ``is a'' relationship with the ``Medical Device'' type, and has a ``contains'' relationship with the ``Clinical Drug'' type.

MEDLINE, the metathesaurus, and the semantic network are represented in our network as different layers. Articles which contain full text abstracts are represented as the abstract layer nodes $\AB$, keywords from the metathesaurus are represented as nodes in the keyword layer $\KW$, and items from the semantic network are represented as nodes in the semantic layer $\SM$.

\subsection{Network Topology}

We define a weighted undirected graph underlying our network $\NET$ as ${G = (V,E)}$, where $V = \AB \cup \KW \cup \SM$.
The construction of $G$ was governed by two major goals. Firstly, the shortest path between two indirectly related keywords should likely contain a significant number of nodes in $\AB$. If instead, this shortest path contained only $\KW-\KW$ edges, we would limit ourselves to known information contained within the UMLS metathesaurus. Secondly, conceptual distance between topics should be represented as the distance between two nodes in $\NET$. This implies that we can determine the similarity between $i,j\in V$ by the weight of their shortest path. If $ij\in E$, this would imply that exists a previously known relationship between $i$ and $j$. We are instead interested in connections between distant nodes, as these potentially represent unknown information. 
Below we describe the construction of each layer in $\NET$.


\begin{figure}[ht]
\centering
\includegraphics[width=\linewidth]{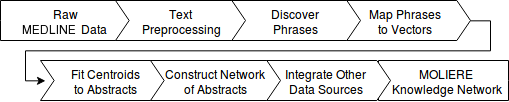}
\caption{\sysname network construction pipeline.}
\label{fig:dataCleaning}
\end{figure}

\begin{figure}[ht]
\centering
\includegraphics[width=0.85\linewidth]{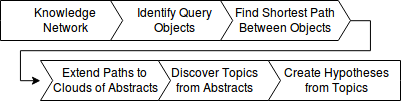}
\caption{\sysname query pipeline.}
\label{fig:queryPipeline}
\end{figure}

\subsection{Abstract Layer $\AB$} \label{articleLayer}

When connecting abstracts ($\AB-\AB$ edges), we want to ensure that two nodes $i,j\in \AB$ with similar content are likely neighbors in the $\AB$ layer. In order to do this, we turned to the UMLS SPECIALIST NLP toolset \cite{System2006} as well as \tmine \cite{el2014scalable} and \ftext \cite{Bojanowski2016,joulin2016bag}. Our process for constructing $\AB$ is summarized in Figure \ref{fig:dataCleaning}.

First, we extract all titles, abstracts, and associated document ID (referred to as PMID within MEDLINE) from the raw MEDLINE files. We then process these combined titles and abstracts with the SPECIALIST NLP toolset to standardize spelling, strip stop words, convert to ASCII, and perform a number of other data cleaning processes. We then use \tmine to generate meaningful $n$-grams and further clean the text. This process finds tokens that appear frequently together, such as \textit{newborn} and \textit{infants} and combines them into a single token \textit{newborn\_infants}. Cleaning and combining tokens in this manner greatly increases the performance of \ftext, the next tool in our pipeline. 

When running \tmine, we keep the minimum phrase frequency and the maximum number of words per phrase set to their default values. We also keep the topic modeling component disabled. On our available hardware, the MEDLINE data set can be processed in approximately thirteen hours without topic modeling, but does not finish within three days if topic modeling is enabled. Because the resulting phrases are of high quality even without the topic modeling component, we accept this quality vs. time trade off. It is also important to note that we modify the version of \tmine distributed by El-Kishky in \cite{el2014scalable} to allow phrases containing numbers, such as gene names like p53.

Next, \ftext maps each token in our corpus to a vector $ v \in \RR^{d}$, allowing us to fit a centroid per abstract $i\in \AB$. Using a sufficiently high-dimensional space ensures a good separation between vectors. In other words, each abstract $i\in \AB$ is represented in $\RR^d$ as $c_i = 1/k \cdot \sum_{j=1}^k x_j$, where $x_j$ are \ftext vectors of $k$ keywords in $i$. 

We choose to use the skipgram model to train \ftext and reduce the minimum word count to zero. Because our data preprocessing and \tmine have already stripped low support words, we accept that any n-gram seen by \ftext is important. Following examples presented in \cite{mikolov2013efficient,mikolov2013distributed,joulin2016bag} and others, we set the dimensionality of our vector space $d$ to 500. This is consistent with published examples of similar size, for example the Google news corpus processed in \cite{mikolov2013efficient}. Lastly, we increase the word neighborhood and number of possible sub-words from five to eight in order to increase data quality.
%

Finally, we used FLANN \cite{muja2014scalable} to create nearest neighbors graph from all $i\in \AB$ in order to establish $\AB-\AB$ edges in $E$. This requires that we presuppose a number of expected nearest neighbors per abstract $k$. We set this tunable parameter to ten initially and noticed that this value seemed appropriate. By studying the distances between connected abstracts, we observed that most abstracts had a range of very close and relatively far ``nearest neighbors". For our purposes in these initial experiments, we kept $k=10$ and saw promising results. Due to time and resource limitations, we were unable to explore higher values of $k$ in this study, but we are currently planning experiments where $k=100$ and $k=1000$. It is important to note that the resulting network will have $\approx k( 2.3\times 10^7)$ edges, so there is a considerable trade-off between quality vs. space and time complexity. 

After experimenting with both $L_2$ and normalized cosine distances, we observed that $L_2$ distance metric performs significantly better for establishing connections between centroids. 
Unfortunately, we cannot utilize the k-tree optimization in FLANN along with non-normalized cosine distance, making it computationally infeasible a dataset of our size. This is because the k-tree optimization requires an agglomerative distance metric. Lastly, we scale edges to the $[0,1]$ interval in order to relate them to other edges within the network.



\subsection{Keyword Layer $\KW$} \label{keyword layer}

The $\KW$ layer is imported from the UMLS metathesaurus. Each keyword is referenced by a CUI number of UMLS. 
This layer links keywords which share already known connections. These known connections are $\KW-\KW$ edges. The metathesaurus connections link related words; for example, the keyword ``Protine p53" \textit{C0080055} is related to ``Tumor Suppressor Proteins'' \textit{C0597611} and ``Li-Fraumeni Syndrome'' \textit{C0085390} among others. There exist 14 different types of connections between keywords representing relationships such as \textit{parent - child} or \textit{broader concept - narrower concept}. We assign each a weight in the $[0,1]$ interval corresponding to its relevance, and then scale all weights by a constant factor $\sigma$ so the average $\AB-\AB$ edge are is stronger than the average $\KW-\KW$ edge. The result is that a path between two indirectly related concepts will more likely include a number of abstracts. We selected $\sigma = 2$, but more study is needed to determine the appropriate edge weights within the keyword layer.

\subsection{$\AB-\KW$ Connections} \label{a2kConnections}

In order to create edges between $\AB$ and $\KW$, we used a simple metric of term frequency-inverse document frequency (tf-idf). UMLS provides not only a list of keywords, but all known synonyms for each keyword. For example, the keyword Color \textit{C0009393} has the American spelling, the British spelling, and the pluralization of both defined as synonyms.
Therefore we used the raw text abstracts and titles (before running the SPECIALIST NLP tools) to calculate tf-idf.
In order to quickly count all occurrences of UMLS keywords across all synonyms, we implemented a simple parser. This was especially important because many keywords in UMLS are actually multi-word phrases such as ``Clustered Regularly Interspaced Short Palindromic Repeats" (a.k.a. CRISPR) \textit{C3658200}. 

In order to count these keywords, we construct a parse tree from the set of synonyms. Each node in the tree contains a word, a set of CUIs, and a set of children nodes, with the exception of the root which contains the null string. We build this tree by parsing each synonym word by word. For each word, we either create a new node in the tree, or traverse to an already existing child node. We store each synonym's CUI in the last node in its parse path.
Then, to parse a document, we simply traverse the parse tree. This can be done in parallel over the set of abstracts. For each word in an abstract, we move from the current tree node to a child representing the same word. If none exists, we return to the root node. At each step of this traversal, we record the CUIs present at each visited node. In this manner, we get a count of each CUI present in each abstract.
Our next pass aggregates these counts to discover the total number of usages per keyword across all abstracts.
We calculate tf-idf per keyword per abstract. Because our network's weights represent distance, we take the inverse of tf-idf to find the weight for an $\AB-\KW$ edge. This is done simply by dividing a CUI's count across all abstracts by its count in a particular abstract. By calculating weights this way, abstracts which use a keyword more often will have a \textit{lower} weight, and therefore, a shorter distance. We scale the edge weights to the $[0,\sigma]$ interval so that these edges are comparable to those within the $\AB$ and $\KW$ layers.

\subsection{Semantic Layer $\SM$} \label{sec:semanticLayer}

The UMLS supplies a companion network referred as the semantic network. 
This network consists of semantic types, which are overarching concepts. These ``types" are similar to the function of a ``type" in a programming language. In other words, it is a conceptual entity embodied by instantiations of that type. In the UMLS network, elements of $\KW$ are analogous to the instantiations of semantic types. While there are over two million elements of $\KW$, there are approximately 130 elements in $\SM$. For example, the semantic type Disease or Syndrome \textit{T047} is defined as ``A condition which alters or interferes with a normal process, state, or activity of an organism" \cite{MedicineUS2009}. There are thousands of keywords, such as ``influenza'' \textit{C0021400} that are instances of this type.

The $\SM-\SM$ edges are connected similarly to $\KW-\KW$ edges. The overall structure is hierarchical with ``Event'' \textit{T051} and ``Entity'' \textit{T071} being the most generalized semantic types. Cross cutting connections are also present and can take on approximately fifty different forms. These cross cutting relations also form a hierarchy of relationship types. For example, ``produces'' \textit{T144} is a more specific relation than its parent ``brings about'' \textit{T187}.

We initially included $\SM$ in our network by linking each keyword to its corresponding semantic type. Unfortunately, in our early results we found that many shortest paths traversed through $\SM$ rather than through $\AB$. For example, if we were interested in two diseases, it was possible for the shortest path would simply travel to the ``Disease or Syndrome'' \textit{T047} type. This ultimately degraded the performance of our hypothesis generation system. As a result we removed this layer, but that further study may find that careful choice of $\SM-\SM$ and $\KW-\SM$ connection weights may make $\SM$ more useful. This is further discussed in Section \ref{sec:deploymentChallenges}.

\section{Query Process} \label{sec:queryProcess}


The process of running a query within \sysname is summarized in Figure \ref{fig:queryPipeline}.
Running a query starts with the user selecting two nodes $i,j\in V$ (typically, but not necessarily, $i,j\in \KW$). For example, a query searching for the relationship between ``stem cells'' and ``strokes'' would be input as keyword identifiers \textit{C0038250} and \textit{C1263853}, respectively. This process simplifies our query process, but determining a larger set of keywords and abstracts which best represents a user's search query is a future work direction.

After receiving two query nodes $i$ and $j$, we find a shortest path between them, $(ij)_{\mathsf{s}}$, using Dijkstra's algorithm. These paths typically are between three and five nodes long and contain up to three abstracts (unless the nodes are truly unrelated, see Section \ref{sec:netstat}). We observed that when $(ij)_{\mathsf{s}}$ contains only two or three nodes in $\KW$, that the $ij$ relationship is clearly well studied because it was solely supplied by the UMLS layer $\KW$. We are more interested in paths containing abstracts because these represent keyword pairs whose relationships are less well-defined. Still, the abstracts we find along these shortest paths alone are not likely to be sufficient to generate a hypothesis.

\subsection{Hypothesis Modeling}

\begin{figure}
\centering
\includegraphics[width=0.7\linewidth]{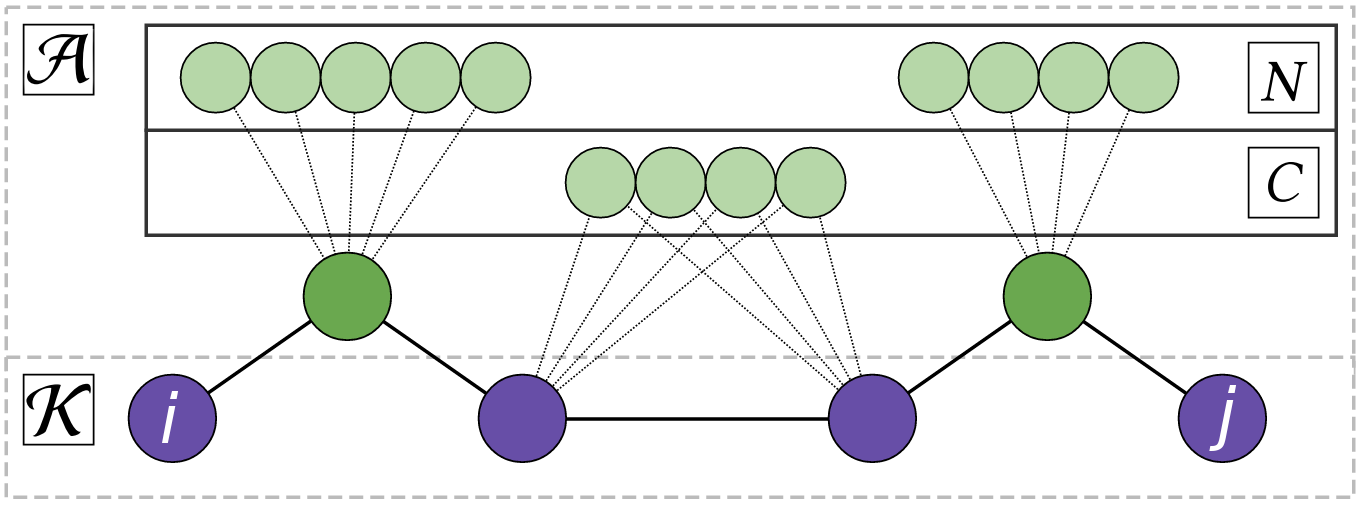}
\caption{Process of extending a path to a cloud of abstracts.}
\label{fig:shortPath}
\end{figure}

Broadening $(ij)_s$ consists of two main phases, the results of which are depicted in Figure \ref{fig:shortPath}. First, we select all nodes $S = (ij)_s \cap \AB$. These abstracts along the path $(ij)_s$ represent papers which hold key information relating two unconnected keywords. We find a neighborhood around $S$ using a weighted breadth-first traversal, selecting the closest 1,000 abstracts to $S$. We will call this set $N$. Because $\AB$ was constructed as a nearest neighbors graph, it is likely that the concepts contained in $N$ will be similar to the concepts contained in $S$, which increases the likelihood that important concepts will be detected by PLDA+ later in the pipeline.

Next, we identify abstracts with contain information pertaining to the $\KW-\KW$ connections present in $(ij)_s$. We do so in order to identify abstracts which likely contain concepts which a human reader could use to understand the known relationship between two connected keywords. We start by traversing $(ij)_s$ to find $\alpha, \beta \in \KW$ such that $\alpha$ and $\beta$ are adjacent in $(ij)_s$. From there, we find a set of abstracts $C=\{c : c\alpha \in E \wedge c\beta \in E\}$. That is, $C$ is a subset of abstracts containing both keywords $\alpha$ and $\beta$. Because $(ij)_s$ can have many edges between keywords, and because thousands of abstracts can contain the same two keywords, it is important to limit the size of $C$.


This process creates a set of around 1,300 $\AB$-nodes. This set will typically contain around 15,000-20,000 words and is large enough for PLDA+ to find topics. We run PLDA+ and request 20 topics. We find this provides a sufficient spread in our resulting data sets. The trained model generated by PLDA+ is what is eventually returned by our query process.

For our experiments, we often must process tens of thousands of results and thus must train topic models quickly. This is most apparent when running a one-to-many query such as the drug repurposing example in \ref{sec:drugrepurposing}. Additionally, the training corpus returned from a \sysname query is often only a couple thousand documents large. As a result, we set the number of topics and the number of iterations to relatively small values, 20 and 100 respectively. Because we store intermediary results, it is trivial to retrain a topic model if the preliminary result seems promising.

The process of analyzing a topic model and uncovering a human interpretable sentence to describe a hypothesis is still a pressing open problem. The process as stated here does have some strong benefits which are apparent in Section \ref{experiments}. These include the ability to find correlations between medical objects, such as between a drug and multiple genes. In Section \ref{sec:lessonsLearned} we explain our initial plans to improve the quality of results which can be deduced from these topic models.

\section{Experiments} \label{experiments}

We conduct two major validation efforts to demonstrate our system's potential for hypothesis generation. For each of these experiments we use the same set of parameters for our trained model and network weights. Our initial findings show our choices, detailed in Section \ref{sec:networkConstruction}, to be robust. We plan to refine these choices with methods described in Section \ref{sec:lessonsLearned}. 

We repeat an experiment done by Wang et al. in \cite{wang2011finding} wherein we discover the implicit connections between the drug Venlafaxine and the genes HTR1A and HTR2A. We also perform a large scale study of Dead Box RNA Helicase 3 (DDX3) and its connection to cancer adhesion and metastasis. Each of these experiments is described in greater detail in the following sections. \emph{In this paper, we deliberately do not evaluate our experiments with extremely popular objects such as p53. These objects are so highly connected within $\KW$ that hypothesis generation involving these keywords is easy for many different methods.}

\subsection{Network Profile}\label{sec:netstat}
We conduct our experiments on a very large knowledge graph which has been constructed according to Section \ref{sec:networkConstruction}. We initially created a network $\NET$ containing information dating up to and including 2016. This network consists of 24,556,689 nodes and 989,169,295 edges. The network overall consists of largest strongly connected component containing 99.8\% of our network. The average degree of a node in $\NET$ is 79.65, and we observe a high clustering coefficient of 0.283. These metrics cause us to expect that the shortest path between two nodes will be very short. Our experiments agree, showing that most shortest paths are between three and six nodes long. 


\subsection{Venlafaxine to HTR1A} \label{sec:venToHTR}

Wang et al. in \cite{wang2011finding} use a similar topic modeling approach, and find during one of their experiments that Venlafaxine \textit{C0078569} appears in the same topic as the HTR1A  and HTR2A genes (\textit{C1415803} and \textit{C1825553} respectively). When looking into these results, they find a stronger association between Venlafaxine and HTR1A. This finding is important because Venlafaxine is used to treat depressive disorder and anxiety, which HTR1A and HTR2A have been thought to affect, but as of 2009 no abstract contains this link. As a result, this implicit connection is difficult to detect with many existing methods.

\begin{figure}
\centering
\includegraphics[width=0.9\linewidth]{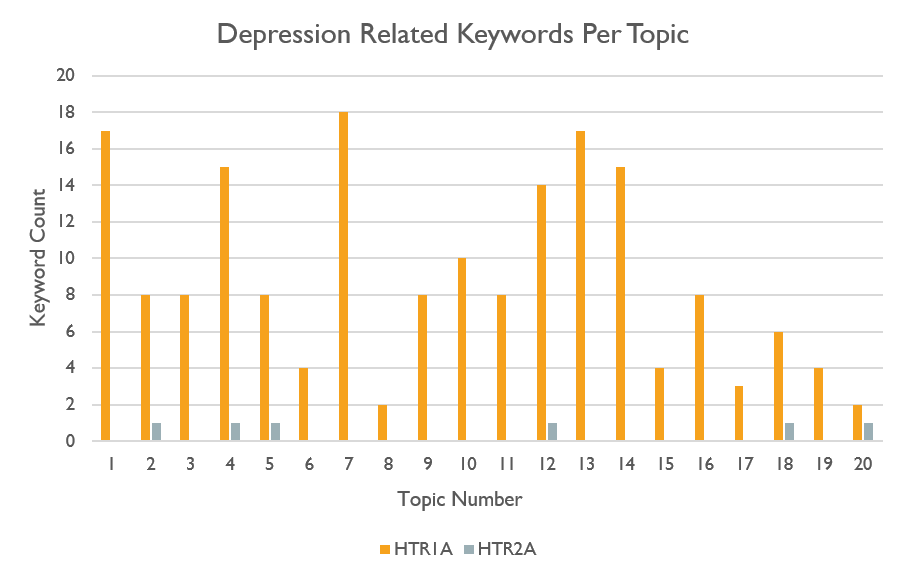}
\caption{Distribution of n-grams having to do with depression from Venlafaxine queries.}
\label{fig:depress}
\end{figure}

\noindent {\bf Results:} As a result of running two queries, Venlafaxine to HTR1A, and Venlafaxine to HTR2A, we can corroborate the findings of Wang et al. in \cite{wang2011finding}. We find that neither pair of keywords is directly connected or connected through a single abstract. Nevertheless, phrases such as ``long term antidepressant treatment,'' ``action antidepressants,'' and ``antidepressant drugs'' are all prominent keywords in the HTR1A query. Meanwhile, the string ``depress'' only occurs four times in unrelated phrases with the HTR2A results. The distribution of depression related keywords from both queries can be see in figure \ref{fig:depress}.

\begin{figure}
\centering
\includegraphics[width=0.9\linewidth]{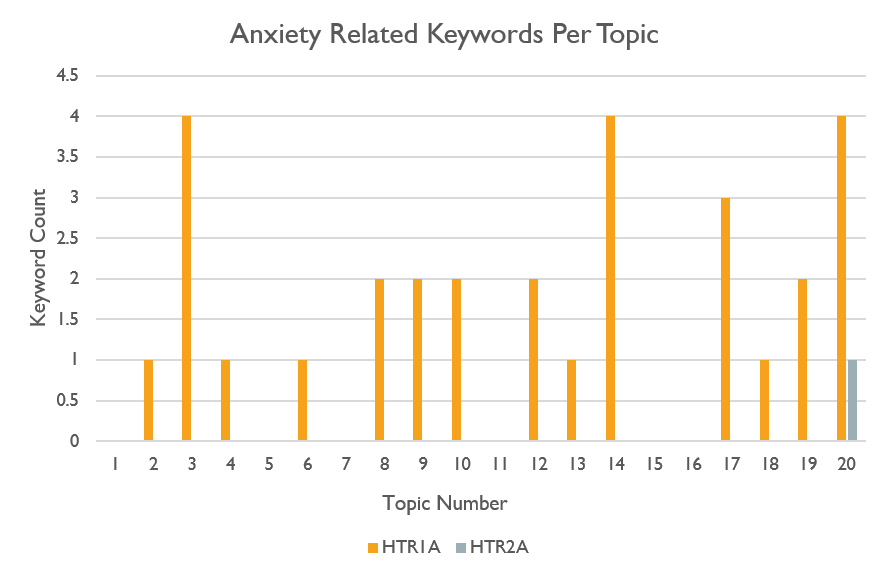}
\caption{Distribution of n-grams having to do with anxiety from Venlafaxine queries.}
\label{fig:anxiety}
\end{figure}

Similarly, our results for HTR1A contain a single topic holding the phrases ``anxiogenic,'' ``anxiety disorders,'' ``depression anxiety disorders,'' and ``anxiolytic response.'' In contrast, our HTR2A results do not contain any phrases related to anxiety. The distribution of anxiety related keywords from both queries can be see in figure \ref{fig:anxiety}.

Our findings agree with those of Wang et al. which were that a small association score of 0.34 between Venlafaxine and HTR1A indicates a connection which is likely related to depressive disorder and anxiety. The association score between Venlafaxine and HTR2A, in contrast, is a much higher 4.0. This indicates that the connection between these two keywords is much weaker.

\subsection{Drug Repurposing and DDX3's Anti-Tumor Applications}
\label{sec:drugrepurposing}

Many genes are active in multiple cellular processes and in many cases they are found to be active outside of the original area in which the gene was initially discovered. The prediction of new processes is especially important for repurposing existing drugs (or drug target genes) to a new application \cite{andronis2011literature, oprea2012drug, allison2012ncats}.  As an example, the drugs developed for the treatment of infectious diseases were recently repurposed for cancer treatment.  Extending applications of existing drugs provides a tremendous opportunity for the development of cost-effective treatments for cancers and other life-threatening diseases. 

To estimate the predictive value of our system for the discovery of new applications of small molecules we select Dead Box RNA Helicase 3 (DDX3) \textit{C2604356}. DDX3 is the member of Dead-box RNA helicase and was initially discovered to be a regulator of transcription and propagation of Human Immunodeficiency Virus (HIV) as well as ribosomal biogenesis.  Initially, DDX3 was a target for the development of anti-viral therapy for the AIDS treatment \cite{kwong2005viral, maga2008pharmacophore}.

More recently, DDX3 activity was found to be involved cancer development and progression mainly through regulation of the Wnt signaling pathway \cite{cruciat2013rna, yim2013unwinding} and associated regulation of Cell-cell and Cell-matrix adhesion, tumor cells invasion, and metastasis \cite{chen2015ddx3, sun2011role, van2016prognostic, krol2015network}.  Currently, DDX3 is an established target for anti-tumor drug development \cite{bol2015targeting, samal2015ketorolac, bol2015ddx3} and represents a case for repurposing target anti-viral drugs into the application area of anti-tumor therapy. 

To test this hypothesis, we analyze the data available on and before 12/31/2009, when no published indication of links in between DDX3 and the Wnt signaling were available. We compare DDX3 to all UMLS keywords containing the text ``signal transduction", ``transcription", ``adhesion", ``cancer", ``development", ``translation", or ``RNA" in their synonym list. This search results in 9,905 keywords over which we query for relationships to DDX3. From this large set of results we personally analyze a subset of important pairs.



\noindent {\bf Results:} In our generated dataset, we found following text grouping within topics: ``substrate adhesion,'' ``RGD cell adhesion domain,'' ``cell adhesion factor,'' ``focal adhesion kinase'' which are indicative for the cell-matrix adhesion. The topics ``cell-cell adhesion,'' ``regulation of cell-cell adhesion,'' ``cell-adhesion molecules'' indicate the involvement of DDX3 into cell-cell adhesion regulation. The involvement of adhesion is associated with topics related to tumor dissemination: `` Collaborative staging metastasis evaluation Cancer,'' ``metastasis adhesion protein, human,'' ``metastasis associated in colon cancer 1'' (selected in between others similar topics). 
    
	The results above suggested that through analysis of the $\leq$2009 dataset we can predict the involvement of DDX3  in tumor cell dissemination through the effects of Cell-cell and cell-matrix adhesion. Next, we analyzed, whether it will be possible to made inside of the mechanisms of DDX3-dependent regulation of Wnt signaling. As shown recently, DDX3 involvement on Wnt signaling is based on the regulated Casein kinase epsilon, to affect phosphorylation of the disheveled protein. Although we cannot predict the exact mechanism of DDX3 based on $\leq$2009 dataset, the existence of multiple topics of signal-transduction associated kinases, like ``CELL ADHESION KINASE'', ``activation by organism of defense-related host MAP kinase-mediated signal transduction pathway'', ``modulation of defense-related symbiont mitogen-activated protein kinase-mediated signal transduction pathway by organism'', suggested the ability of DDX3 to regulate kinases activities and kinase-regulated  pathways. 
    
\subsection{Experimental Setup}
\label{sec:expirementalSetup}

We performed all experiments on a single node within Clemson's Palmetto supercomputing cluster. To perform our experiments and construct our network, we use an HP DL580 containing four Intel Xeon x7542 chips. This 24 core node has 500 GB of memory and access to a large ZFS-based file system where we stored experimental data. 

For the DDX3 queries, we initially searched for all $(ij)_s$ where $i=\text{DDX3}$ and $j \in \KW$. This resulted in 1,350,484 shortest paths with corresponding abstract clouds. We used PLDA+ to construct models for all of these paths. Discovering all $(ij)_s$ completed in almost 10 hours of CPU time, and training the respective models completed in slightly over 68 hours of CPU time. We ran PLDA+ in parallel, resulting in a wall time of only 12 hours. As mentioned previously, this large dataset was filtered to the 9,905 paths we are interested in.


We generate the results for the Venlafaxine experiments in one hour of CPU time, which is mostly spent loading our very large network and then running Dijkstra's algorithm. After this, the two resulting PLDA+ models were trained in parallel within a minute.

\section{Deployment Challenges} \label{sec:deploymentChallenges}

In the following section we detail the challenges which we have faced and are expecting to encounter while creating our system and deploying it to the research community.

\noindent\textbf{Dynamic Information Updates} The process of creating our network is computationally expensive and for the purposes of validation we must create multiple instances of our network representing different points in time. Initially we would have liked to create these multiple instances from scratch, starting from the MEDLINE archival distribution and rebuilding the network from there. Unfortunately, this proved infeasible because creating a single network is a time consuming process. Instead, we filter our network by removing abstracts and keywords which were published after our select date. Additionally, the act of adding information to our network, such as extending the 2016 network to create a 2017 network, is not straightforward. Ideally, adding a small number abstracts or keywords should be a fast and dynamic process which only affects localized regions of the network. If this were so, our deployed system could take advantage of new ideas and connections as soon as they are published.

A deployed system could support dynamic updates with an amortized approach. Using previously created \ftext and \tmine models, new documents could be fitted into an existing network with suitably high performance. Of course, if a new document introduced a new keyword or phrase, we would be unable to detect it initially. After some threshold of new documents had been added to the network, we could then rerun the entire network construction process to ensure that new keywords, phrases, and concepts would be properly placed in the network.

\noindent\textbf{Query Platform and Performance:}
Initially, we expected to use a graph database to make the query process easier. We surveyed a selection of graph databases and found that Neo4j \cite{developers2012neo4j} provides a powerful query language as well as a platform capable of holding our billion-edge network. Unfortunately, Neo4j does not easily support weighted shortest path queries. Although some user suggestions did hint that it may be possible, the process requires leveraging edge labels and custom java procedures in a way that did not seem scalable. In place of Neo4j, we implemented Dijkstra's shortest path algorithm in C++ using skew heaps as the internal priority queue. This implementation was chosen to minimize memory usage while maximizing speed and readability. Because we implemented Dijkstra's algorithm ourselves, we can also combine the process of finding a shortest path and finding all neighboring abstracts for all keywords from a specific source. With only these high level optimizations, we were able to generate over 1,350,000 shortest paths and abstract neighborhoods in under ten hours, but generating a single result takes slightly over one hour.

\section{Lessons Learned and Open Problems} \label{sec:lessonsLearned}

\noindent {\bf Specialized LDA:} During last two decades there has been a number of significant attempts to design automatic hypothesis generation systems \cite{spangler2015accelerating,swanson1998link,wang2011finding}. However, most of these improve their performance by restricting either their information space or the size of their dictionary. For example, specialized versions of LDA such as Bio-LDA \cite{wang2011finding} uncover latent topics using a dictionary that gives a priority to special terms. We find that such approaches are helpful when general language may significantly over weigh a specialized language. However, phrase mining approaches that recover $n$-grams, such as \cite{el2014scalable}, produce accurate methods without limiting the dictionary.

\noindent {\bf Hypothesis Viability and Novelty Assessment:} Intuitively, a strong connection between two concepts in $\NET$ means that there exist a significant amount of research that covers a path between them. Similar observations are valid for LDA, i.e., latent topics are likely to describe well known facts. As a result, the most meaningful connections and interpretable topical inference are discovered with latent keywords that are among the most well known concepts. However,
real hypotheses are not necessarily described using the
most latent keywords in such topic models. In many cases, the
keywords required for a successful and interpretable hypothesis
start to appear among 20-30 most latent topical keywords. 
Thus, a major open problem related is the process to which one should select a combination of keywords and topics in order to represent a viable hypothesis. This problem is also linked to the problem of assessing the viability of a generated hypothesis.

These problems, as well as the problem of hypothesis novelty assessment, can be partially addressed by using the Dynamic Topic Modeling (DTM) \cite{blei2006dynamic}. Our preliminary experiments with scalable time-dependent clustered LDA \cite{gropp2016scalable} that significantly accelerates DTM demonstrate a potential to discover dynamic topics in MEDLINE. The dynamic topics are typically more realistic than those that can be discovered in the static network. This significantly simplifies the assessment of viability and topic noise elimination.

\noindent\textbf{Incorporating the Semantic Layer $\SM$:}
In section \ref{sec:semanticLayer} we describe the process in which we evaluated the UMLS semantic network and found that it worsened our resulting shortest path queries. Further work could improve the contribution that $\SM$ has on our overall network, possibly allowing $\SM$ to define the overall structure of our knowledge graph. In order to do this, one would likely need to take into account the hierarchy of relationship types present in this network, as well as the relative relationship each element in $\KW$ has with its connection in $\SM$. Ultimately, these different relationships would need to inform a weighting scheme that balances the over generalizations that $\SM$ introduces. For example, it may be useful to understand that two keywords are both diseases, but it is much less useful to understand that two keywords are ``entities".

\noindent {\bf Learning the Models of Hypothesis Generation:} 
There is surprisingly little research focused on addressing the process of biomedical research and how that process evolves over time. We would like to model the process of discovery formation, taking into account the information context surrounding and preceding a discovery. We believe we could do so by reverse engineering existing discoveries in order to discover factors which altered the steps in a scientist's research pipeline. Several promising observations in this direction have been done by Foster et al. \cite{foster2015tradition} who examined this through Bourdieu's field theory of science by developing a typology of research strategies on networks extracted from MEDLINE. However, instead of reverse engineering their models, they separate innovation steps from those that are more traditional in the research pipeline.


\noindent \textbf{Dynamic Keyword Discovery:} One of the limitations we found when performing our historical queries is the delay between the first major uses of a keyword and its appearance in the UMLS metathesaurus. Initially, we planned to study the relationship between ``CRISPR'' \textit{C3658200} and ``genome editing'' \textit{C4279981}. To our surprise, many keywords related to this query did not exist in our historical networks between 2009 and 2012, despite their frequent usage in cutting-edge research during that time. To further confuse the issue, although the keyword ``CRISPR'' did not appear in the UMLS releases on or before 2012, keywords containing ``CRISPR'' as a substring, such as ``CRISPR element metabolism'' \textit{C1752766}, do appear. We find this to be contradictory and that these inconsistencies highlight the limitation of relying on so strongly on keyword databases. Going forward, we plan to devise a way to extend a provided keyword network, utilizing semantic connections we can find within the MEDLINE document set. Projects like \cite{spangler2014automated} have already shown this method can work in domains of smaller scales with good results. The challenge will be to extend this method to perform well when used on the entire MEDLINE data set.

\noindent \textbf{Improving Performance of Algorithms with Graph Reordering Techniques:} Cache-friendly layouts of graphs are known to generally accelerate the performance of the path and abstract retrieval algorithms which we apply. Moreover, it is desirable to consider this type of acceleration in order to make our system more suitable for regular modern desktops. This is an important consideration as memory is not expected to be a major bottleneck after the network is constructed. We propose to rearrange the network nodes by minimizing such objectives as the minimum logarithmic or linear arrangements \cite{SafroT11,SafroRB06}. On a mixture of $\KW-\KW$, $\AB-\KW$, and $\AB-\AB$ edges we anticipate an improvement of  at least 20\% in the number of cache misses according to \cite{safro2009improving}.

\noindent \textbf{Mass Evaluation:} We note that evaluation techniques are largely an issue in the state of the art of hypothesis generation. While some works feature large scale evaluation performed by many human experts, a majority, this work included, are restricted to only a couple of promising results to justify the system. In order to better evaluate and compare hypothesis generation techniques we must devise a common and large scale suite of historical hypotheses. We are currently evaluating whether a ground-truth network, like the drug-side-effect network SIDER \cite{lee2011building}, can be a good source of such hypotheses. For example, if we identify a set of recently added connections within SIDER, and predict a substantial percentage of those connections using \sysname, then we may be more certain of our performance.

\noindent \textbf{New Domains of Interest:} We have considered other domains on which \sysname may perform well. These include generating hypotheses regarding economics, patents, narrative fiction, and social interactions. These are all domains where a hypothesis would involve finding new relationships between distinct entities. We contrast this with domains such as mathematics where the entity-relationship network is much less clear, and logical approaches from the field of automatic theorem proving are more applicable.


\section{Conclusions} \label{conclusion}
In this study we describe a deployed biomedical hypothesis generation system, \sysname, that can discover relationship hypotheses among biomedical objects. This system utilizes information which exists in MEDLINE and other NLM datasets. We validate \sysname on landmark discoveries using carefully filtered historical data. Unlike several other hypothesis generation systems, we do not restrict the information retrieval domain to a specific language or a subset of scientific papers since this method can lose an unpredictable amount of information. Instead, we use recent text mining techniques that allow us to work with the full heterogeneous data at scale. We demonstrate that \sysname successfully generates hypotheses and recommend using it to advance biomedical knowledge discovery. Going forward, we note a number of directions along which we can improve \sysname as well as many existing hypothesis generation systems. 

\section{Acknowledgments} \label{acknowledgments}
We would like to thank Dr. Lihn Ngo for his help in using the Palmetto supercomputer which ran our experiments, and Cong Qiu for initial experiments with Neo4j.

\bibliographystyle{plain}
\bibliography{sample}

\end{document}